# Curvature-driven revival of charge density waves in non-Euclidean space

Zhipeng Song[1,2], Zeyu Liu[3], Junde Liu[1,2], Yi Biao[1,2], Anning Yang[1,2], Qian Fang[1,2], Mojun Pan[1,2], Chen Liu[4], Jiaou Wang[4],

Tian Qian[1,2], Chenmin shen[1,2], Hongliang Lu[1,2]✉, Wei Ji[3]✉, Hong-Jun Gao[1,2], Xiao Lin[1,2]✉

[1]*School of Physical Sciences, University of Chinese Academy of Sciences, Chinese Academy of Sciences, Beijing 101408, China*

[2]*Institute of Physics and Beijing National Center for Condensed Matter Physics, Chinese Academy of Sciences, Beijing 100190, China.*

[3]*Beijing Key Laboratory of Optoelectronic Functional Materials & Micro-Nano Devices, Department of Physics, Renmin University of China, Beijing 100872, China*

[4]*Institute of High Energy Physics, Chinese Academy of Sciences, Beijing 100049, China.*

These authors contributed equally: Zhipeng Song, Zeyu Liu, Junde Liu, Yi Biao

E-mail: luhl@ucas.ac.cn; wji@ruc.edu.cn; xlin@ucas.ac.cn

Strongly correlated quantum states, such as charge density waves (CDWs), are exquisitely sensitive to Fermi surface topology and lattice symmetry, and are typically quenched by heavy carrier doping. In two-dimensional (2D) systems, however, macroscopic geometric curvature emerges as a novel structural degree of freedom to modulate microscopic quantum coherence. This raises a compelling physical question: can non-Euclidean geometric deformations compete with extreme electronic perturbations to reshape, or even revive, a quenched macroscopic quantum order? Here, by constructing monolayer $TiSe_2$-$NbSe_2$ heterostructure on a BLG/SiC substrate for the first time, we report the curvature-driven revival of a frustrated charge order in a non-Euclidean space. Low-temperature angle-resolved photoemission spectroscopy (ARPES) reveals a massive interfacial charge transfer, which destroys the global Fermi surface nesting and completely suppresses the long-range CDW order in Euclidean flat regions. Strikingly, high-resolution scanning tunneling microscopy (STM) reveals that a novel, non-linear CDW state miraculously survives, remaining strictly localized within morphologically distorted, non-Euclidean nanoscale curved regions. Atomistic simulations unravel the structural origin of this phenomenon, demonstrating that interfacial twist and lattice mismatch spontaneously generate a corrugated superlattice. The resulting interplay between stacking-dependent interlayer coupling and local non-Euclidean curvature deterministically dictates the fate of the CDW, suppressing it in flat regions while preserving and geometrically distorting it within highly curved domains. Our findings demonstrate that macroscopic non-Euclidean geometry can serve as a powerful gauge-field tool to rescue and sculpt strongly correlated quantum states in hostile electronic environments, establishing a new paradigm for manipulating topological quantum phases.

Charge density waves (CDWs) are typical collective quantum phases in condensed matter physics, arising from electronic instabilities in low-dimensional metallic systems [1-4]. Over the past few decades, experimental systems [5-11] and theoretical frameworks [12-20] for CDWs have been primarily based on flat Euclidean space. Unlike traditional Euclidean space, non-Euclidean geometries, such as hyperbolic lattices and curved spaces, provide a new dimension for electronic behaviors. This is crucial for revealing the intrinsic coupling between macroscopic quantum order and real-space geometric curvature and it also facilitates a deeper understanding of how geometry influences correlated electronic behaviors.

In non-Euclidean spaces, CDWs exhibit physical properties that are distinct from those in flat spaces. First, geometric frustration plays a critical role: due to the absence of global translational symmetry in curved space, the phase coherence of the CDW is strongly modulated by geometric curvature [21, 22]. Second, non-Euclidean geometry alters the local metric of the system. The curvature effect is equivalent to the introduction of a non-uniform gauge field into the Hamiltonian, which not only induces distortion of the Fermi surface structure but also leads to a spatially localized distribution of electron-phonon coupling strength [23, 24]. In flat space, CDWs typically feature a globally uniform wave vector **q** [1, 5, 6, 11, 25, 26]. In contrast, within non-Euclidean space, **q** evolves with spatial curvature, resulting in spatially dependent modulation patterns. Although non-Euclidean geometry provides rich and exotic physical properties, significant experimental challenges remain. It is difficult to precisely construct 2D materials with non-Euclidean geometries at the nanoscale to study their CDW behaviors. Furthermore, in flat Euclidean space, the superstructure period of conventional CDWs (such as $TiSe_2$) exhibits linear characteristics. Specifically, the superstructure phase follows a linear trajectory [19, 27]. This raises a critical question: Can non-linear CDW states emerge in non-Euclidean spaces?

In this work, we report the observation of a CDW residing in a non-Euclidean space within a monolayer 1T-$TiSe_2$/2H-$NbSe_2$ van der Waals heterostructure epitaxially grown on a bilayer graphene (BLG)/SiC substrate. The successful fabrication of this novel two-dimensional architecture is unambiguously verified by combining scanning tunneling microscopy (STM) and X-ray photoelectron spectroscopy (XPS). Notably, atomic-resolution STM imaging reveals a unique non-Euclidean periodic modulation. The corresponding fast Fourier transform (FFT) analysis exhibits a multi-oriented 2×2 superstructure, breaking the translational symmetry of the uniform 2×2 linear superlattice conventionally observed in pristine 1T-$TiSe_2$. In contrast, this superstructure is not observed in room-temperature STM images. Furthermore, scanning tunneling spectroscopy (STS) and angle-resolved photoemission spectroscopy (ARPES) consistently indicate a strong electron-doping effect, which induces an indirect bandgap below the Fermi level. Our computational simulations reveal that twist-induced structural corrugations deterministically control the CDW state, suppressing it in flat regions via interlayer coupling while preserving and distorting it within highly curved domains. These results unveil an unprecedented non-Euclidean spatial CDW state in 1T-$TiSe_2$.

The schematic of the fabricated heterostructure is depicted in Fig. 1a, showing a stacked structure with a monolayer $TiSe_2$ on top, a monolayer $NbSe_2$ in the middle, and a BLG/SiC(0001) substrate at the bottom. In Fig. 1b, a representative STM image is presented, showing the morphological image of the monolayer $TiSe_2$ on monolayer 2H-$NbSe_2$. The thickness of the monolayer $TiSe_2$ is approximately 0.56 nm, as shown in the inset of Fig. 1b. Figure 1c shows an atomically resolved STM image acquired at 4.5 K on a flat region of the heterostructure surface. Notably, no CDW is observed in the flat $TiSe_2$ regions. This absence is attributed to strong electron doping that completely suppresses the CDW state, as elucidated by the subsequent STS and ARPES data. To further confirm the heterostructure and rule out the presence of other compounds, we employed XPS to analyze the chemical compositions of the sample. The XPS peaks in Fig. S2a at 461.5 and 455.4 eV are indicative of the Ti $2p_{1/2}$ and $2p_{3/2}$ states in $TiSe_2$, respectively[28], while in Fig. S2b, the XPS peaks at 206.1 eV and 203.2 eV respectively represent the Nb $3d_{3/2}$ and Nb $3d_{5/2}$ states in 2H-$NbSe_2$[29]. In Fig. S2c, the peaks at 54.4 and 53.4 eV match the Se $3d_{3/2}$ and $3d_{5/2}$ core levels in 1T-$TiSe_2$. Additionally, peaks at 54.9 and 53.9 eV represent Se $3d_{3/2}$ and $3d_{5/2}$ in 2H-$NbSe_2$. Combining STM and XPS characterizations, we confirm the formation of monolayer 1T-$TiSe_2$/2H-$NbSe_2$ heterostructure.

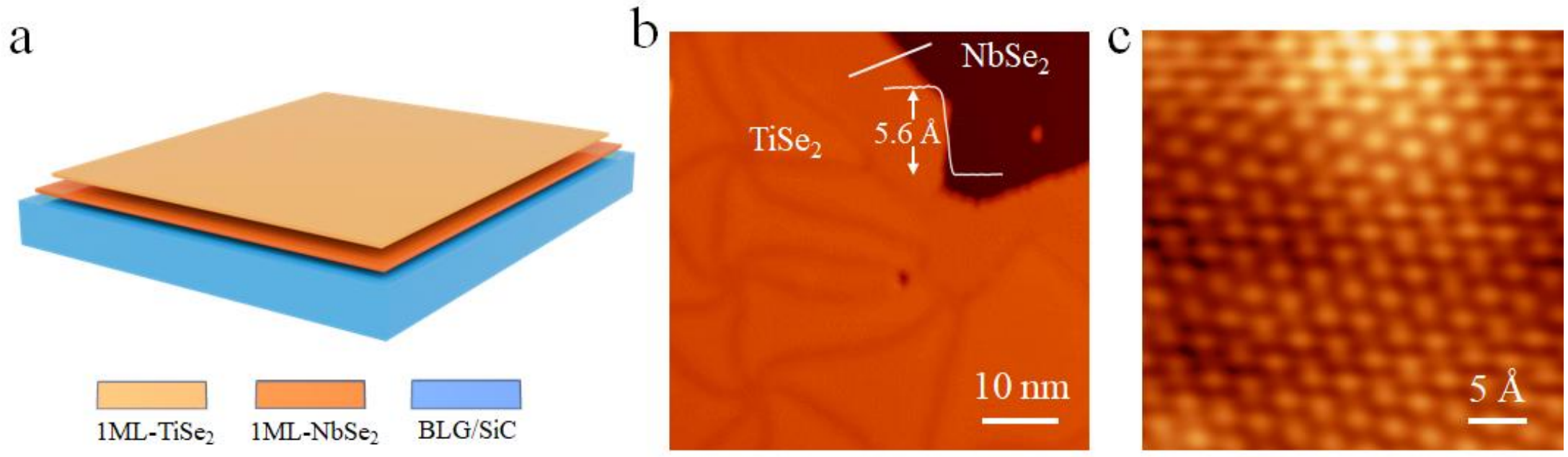

**Fig.1 | The formation of monolayer $TiSe_2/NbSe_2$ heterostructure on a BLG/SiC(0001) substrate. a,** Schematic illustration of the heterostructure of monolayer $TiSe_2/NbSe_2$ on BLG. **b,** Large-scale STM image of monolayer $TiSe_2$ grown on monolayer $NbSe_2$ (V=−2 V, I=200 pA, T=4.5 K). **c,** Atomic-resolution STM image of the monolayer $TiSe_2$ at the flat surface (V=−0.5 V, I=100 pA, T=4.5 K).

Monolayer 1T-$TiSe_2$, when grown on the surface of graphene, is atomically flat and exhibits a 2×2 superstructure, as depicted in Fig. S1d. The unique morphological characteristics emerge when monolayer 2H-$NbSe_2$ is initially grown on graphene, followed by the growth of monolayer 1T-$TiSe_2$ on the $NbSe_2$ surface. This distinct morphology is illustrated in Fig. 2a, where six curved domain boundaries converge to form a vortex-like topology. Within this structural figure, superstructures can be observed in some of the domains especially in their central areas and the boundaries. Notably, some areas lack such superstructures. Figure 2b provides the atomic-resolution STM image of monolayer vortex-like $TiSe_2$（V-$TiSe_2$）in the blue square of Fig. 2a, revealing the presence of nonlinear superstructures within this specified area. To understand the relationship between these superstructures and the intrinsic structure of V-$TiSe_2$, we performed a FFT of the STM image. As displayed in Fig. 2e, the outer six points represent the intrinsic structure of V-$TiSe_2$ while the middle circle indicates the 2×2 superstructure, suggesting its presence in various directions. 1T-$TiSe_2$, a transition metal dichalcogenide, is a fascinating 2D layered material that displays a commensurate (2×2×2) superstructure in bulk below a transition temperature of approximately 200 K[30], and this superstructure is absent at room temperature. Similarly, for the V-$TiSe_2$ we synthesized, a nonlinear 2×2 superstructure appears. To further confirm this as a CDW, we conducted room temperature STM characterization, as shown in Figure S3a. In this figure, a hexapetalous vortex-like morphology can be observed. Unlike at low temperatures, the room temperature STM image lacks the 2×2 superstructure, as indicated by the Fourier transform in Fig. S3b, revealing the characteristics of the absence of the superstructure. Therefore, we conclude that the nonlinear superstructure observed on the surface of V-$TiSe_2$ is a new CDW type.

In the vortex-like morphology shown in Fig. 2a, a spatially alternating phase distribution can be identified: the α region exhibiting a nonlinear 2 × 2 CDW superstructure, whereas the β region retaining the unreconstructed 1 × 1 lattice. The β region do not exhibit superstructures, as illustrated in Fig. 2c and its Fourier transform in Fig. 2f. This suggests that the nonlinear CDW appears only in localized special morphological structures, not across the entire sample surface. To investigate the microscopic origin of this local phase separation, we extracted representative height profiles in Fig. 2d, where the blue and yellow curves correspond to the arrowed paths in the α and β regions of Fig. 2a, respectively. Comparative analysis shows that the local deformation curvature in the α region ($\kappa_{\text{blue}} \approx 0.0631\ \text{nm}^{-1}$) is significantly larger than that in the β region ($\kappa_{\text{yellow}} \approx 0.0479\ \text{nm}^{-1}$) (see the Supplementary Materials for the calculation of Curvature). This difference in surface geometry is precisely the origin of the emergent anomalous CDW state in the α region.

To gain a clearer understanding of the detailed characteristics of this nonlinear superstructure, we enlarged the position of the blue square in Fig. 2b to get more clear distinctions of regions with superstructures, as shown in Fig. 2g. The details of the images reveal that these superstructures are in shape. Figure 2i displays the profile along the blue line in Fig. 2g. The profile indicates that the depth of the superstructure's concavity is approximately 15 pm, which is significantly different from the depth of atomic defects. Line profiles extracted along the arrows across the curved region and domain boundary (Fig. S4a) are displayed in Fig. S4b. The black curve indicates a domain boundary depth of ~120 pm, while the blue curve resolves the atomic and superstructure corrugations of V-$TiSe_2$. The notably large depth of the boundary implies a drastic local lattice reconstruction compared to the intrinsic superstructure. As a comparison, the STM image of the conventional CDW of 1T-$TiSe_2$ exhibits a linear and protruding superstructure, as shown in Fig. 2h. Figure 2j shows the profile along the blue line in Fig. 2h, demonstrating that the protrusion height of the CDW superstructure in 1T-$TiSe_2$ is about 50 pm. The blue and gray lines in Fig. 2k represent the representative scanning tunneling spectroscopy (STS) point spectra of monolayer V-$TiSe_2$ and monolayer 1T-$TiSe_2$ on graphene, respectively. There is a slight difference between the two spectra, with the CDW gap of V-$TiSe_2$ located below the Fermi level, similar to the spectrum of $TiSe_2$ doped with Cu atoms.

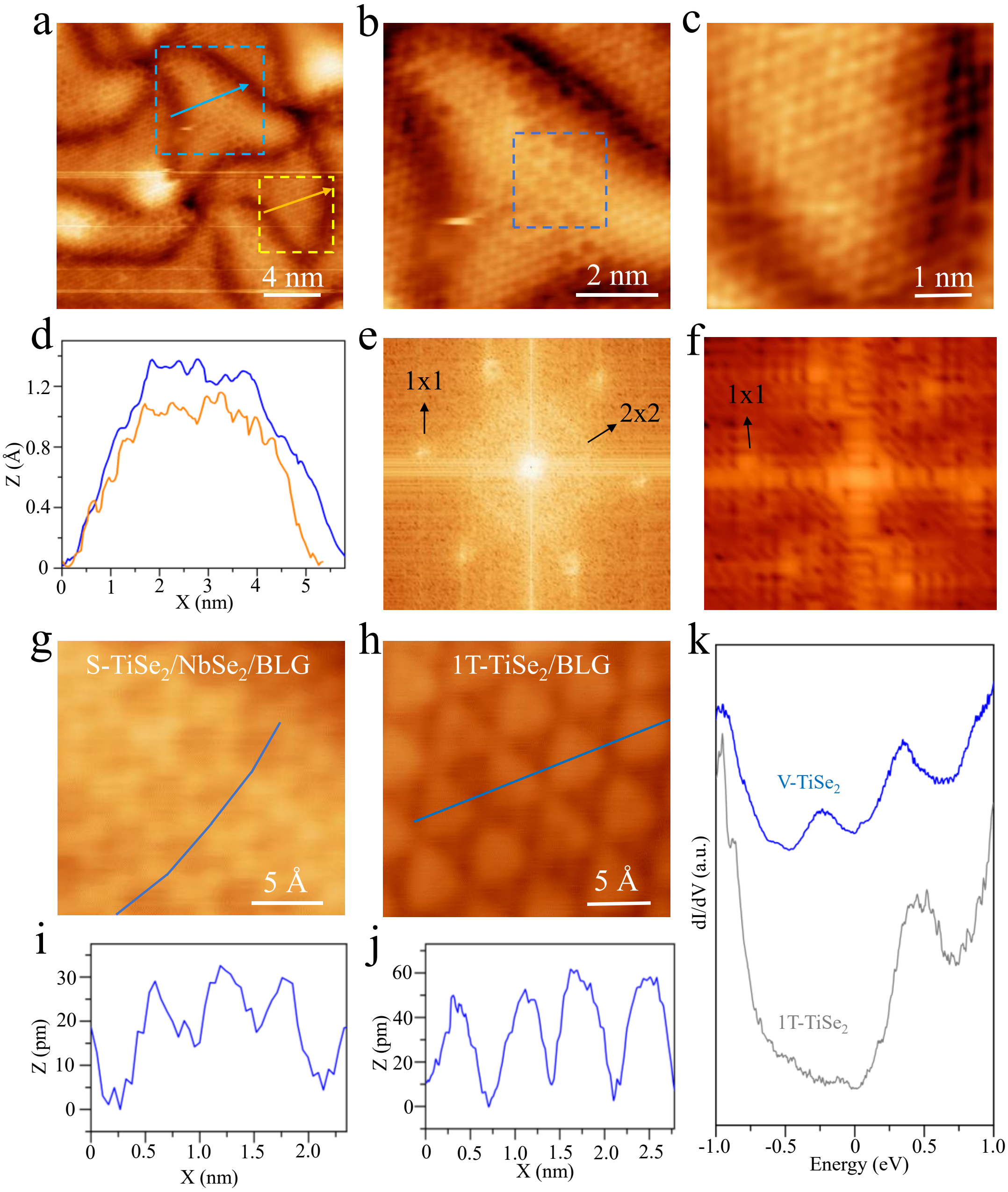

**Fig.2 | Nonlinear charge density wave in $TiSe_2$ with vortex morphology. a,** Large-scale STM image of monolayer V-$TiSe_2$ on the heterostructure, showing a vortex morphology. **b,c** Atomic-resolution STM image of monolayer V-$TiSe_2$ in the blue and yellow square in **a**, respectively.

**b** displays the superstructure, whereas this superstructure is absent in **c**. **d,** Line profile taken along the blue and yellow arrows in a, showing the curvature of the surface. **e,f** Fast Fourier transform (FFT) of STM images in **b** and **c**, respectively. **e** illustrates the 2x2 CDW superstructure in various directions, while **f** does not exhibit this structure. **g,** Small-scale STM image of monolayer V-$TiSe_2$ in the blue square in **b**, showing the nonlinear superstructure. **h,** STM image of monolayer 1T-$TiSe_2$ on graphene, showing the linear superstructure. **i,j** Line profile taken along the blue lines in **g** and **h,** respectively. **k**, Representative STS spectra in **g** and **h**. The blue line represents the STS spectrum of V-$TiSe_2$, while the gray line corresponds to that of 1T-$TiSe_2$ on graphene. All STM images were obtained at V=−0.5 V, I=100 pA and T=4.5 K.

To study the electronic structure of the monolayer V-$TiSe_2$/2H-$NbSe_2$ heterostructure, we employed ARPES for sample characterization at low-temperature condition. The photoemission intensity plot along the Γ-M direction at 6 K is shown in Fig. 3a, where the dispersion bands can be clearly observed. The hole-like bands near the Fermi level at the Γ point are associated with Se 4p bands (Se 1 and Se 2) while the electron-like band near the M point mainly stem from the Ti 3d, consistent with previous ARPES experiments of monolayer 1T-$TiSe_2$/CuSe heterostructure[10]. The energy bands near the Γ point primarily stem from the contribution of 1T-$TiSe_2$[31], while 2H-$NbSe_2$ makes almost no contribution[32]. The red curve of Fig. 3c shows the energy distribution curve (EDC) at the Γ point (Fig.3b). The energy position at the top of the valence band is marked by the dashed line, being 260 meV lower than the Fermi level. The slight difference from the monolayer 1T-$TiSe_2$ on BLG[33] is that the energy band of Se 4P in this heterostructure has shifted downward by an additional 93 meV. It is noteworthy that no folding bands are observed. This is partly due to the inherent weakness of the folding bands generated during CDW occurrences. Additionally, a significant reason is that the CDW superstructures produced by the monolayer V-$TiSe_2$/2H-$NbSe_2$ heterostructure encompass various orientations, greatly diminishing the already faint bands. As a result, these folding bands are almost invisible in ARPES image.

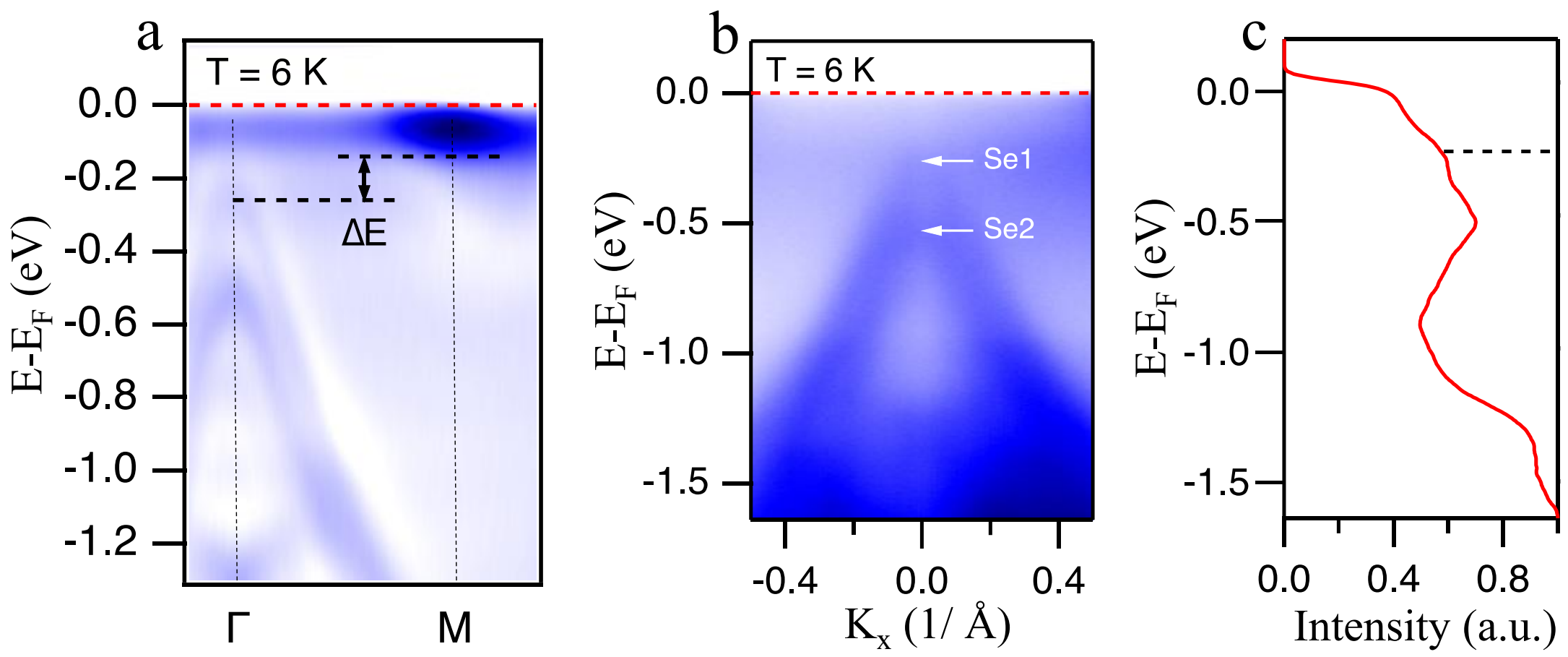

**Fig.3 | Band structures measured at 6 K by ARPES. a**, ARPES intensity plots along Γ-M direction. **b,** Electronic band around Γ point. **c**, Energy distribution curves (EDCs) at Γ point.

It is worth emphasizing, however, that although STM in real space clearly confirms the existence of a nonlinear 2 × 2 CDW superstructure, we do not directly observe folded bands induced by the superlattice potential in ARPES. This apparent discrepancy between momentum-space and real-space signatures in fact reveals the distinctive microscopic nature of the CDW in non-Euclidean space. Under strong electron doping, enhanced long-range Coulomb screening already tends to weaken the folded-band signatures of a CDW driven by either an excitonic-insulator mechanism or electron–phonon coupling. More fundamentally, however, the key physical origin lies in the geometric-topological constraints imposed by the lattice. As demonstrated by the STM results above, the CDW in V-$TiSe_2$ is not a globally extended uniform long-range order, but is instead strictly confined to curved non-Euclidean regions and exhibits strong branching and multiple orientations. In momentum space, such a local superstructure lacking macroscopic long-range translational symmetry necessarily leads to severe momentum broadening of the corresponding reciprocal vectors. This curvature-induced destruction of spatial phase coherence in the quantum state smooths out and ultimately extinguishes the already weak folded-band signal, rendering it invisible in ARPES measurements, which intrinsically average over a macroscopic spatial area.

To clarify the microscopic structural origin of the non-Euclidean CDW modulation observed experimentally, we performed atomistic simulations of a 1.57° twisted $TiSe_2$/$NbSe_2$ heterostructure. In the model, the upper $TiSe_2$ layer is subjected to a 0.75% compressive strain arising from lattice mismatch. As shown in Fig. 4a, the relaxed heterostructure develops a six-lobed vortex-like moiré pattern with a periodicity of approximately 11 nm, as outlined by the red solid lines, consistent with the experimental observations. This reconstruction arises from the competition between intralayer elastic energy and interlayer stacking-dependent interactions. The energetically favored MM stacking and the metastable MX stacking expand into distorted triangular domains, whereas the energetically unfavorable XX stacking region shrinks significantly (Fig. S5). Figure 4b shows the projected spatial distribution of the interlayer Se–Se distance in the relaxed heterostructure. The interlayer-distance map exhibits a vortex-like pattern that follows the moiré superlattice. The strongly coupled MM stacking region has the smallest interlayer separation of approximately 3.0 Å, whereas the weakly coupled XX stacking region shows the largest separation of approximately 3.5 Å. The spatial variation in interlayer coupling further induces pronounced out-of-plane corrugation. As shown in Fig. 4c, the relative height of the topmost Se atoms displays a clear corrugated pattern. Under the C3 symmetry of the moiré lattice, the MM and MX stacking regions bend upward to form six protruding peaks, while the XX regions are depressed downward to form valleys. The topography is consistent with the experimentally observed vortex-like morphology. These results indicate that the 1.57° twisted $TiSe_2/NbSe_2$ interface spontaneously reconstructs into a corrugated moiré superlattice, and this atomic-scale reconstruction further drives the stacking-dependent atomic arrangement within the $TiSe_2$ layer.

Figure 4d presents the smoothed residual Ti atomic density in the $TiSe_2$ layer. The residual density exhibits a well-defined six-lobed vortex-like moiré texture, indicating that the in-plane atomic arrangement of $TiSe_2$ is strongly modulated by the reconstructed moiré geometry. The local Ti density is closely correlated with the stacking configuration, as shown in Fig. 4(e–g). In the MM region, the intrinsic 2×2 CDW distortion of $TiSe_2$ is largely suppressed, presumably due to stronger interlayer coupling, leaving a dominant 1×1 atomic periodicity [Fig. 4e]. In contrast, the MX region retains a clear 2×2 superstructure, indicating that the interlayer coupling there is not strong enough to destroy the intrinsic CDW order [Fig. 4f]. In the XX region, the competition between in-plane elastic energy and interlayer coupling causes the domain area to shrink, resulting in a highly disordered local structure, where no well-defined periodicity can be identified [Fig. 4g].

To further establish the connection between lattice corrugation and atomic-density redistribution, we overlaid the Ti-density residual contours on the local curvature maps of the smoothed Se surface [Fig. 4(h–j)]. The MM region is nearly flat with relatively small curvature [Fig. 4h]. By contrast, the MX region shows much larger curvature that varies anisotropically from the center outward [Fig. 4i]. In this region, the preserved 2×2 superstructure is geometrically distorted by the curved moiré background, giving rise to a nonlinear long-period modulation. The curvature variation in the XX region is also more pronounced than that in the MM region [Fig. 4j], and the Ti-density contours deviate from a simple 1×1 unit-cell pattern, suggesting a possible CDW-like modulation.

These simulations reveal that the interlayer twist and lattice mismatch generate a corrugated, non-Euclidean moiré geometry in the $TiSe_2/NbSe_2$ heterostructure. The resulting stacking-dependent interlayer coupling and local curvature reshape the atomic arrangement of the $TiSe_2$ layer, selectively suppressing the CDW-related 2×2 distortion in the strongly coupled MM regions, preserving it in the MX regions. Therefore, the nonlinear CDW modulation observed experimentally can be understood as a curvature- and stacking-modulated 2×2 lattice distortion confined within specific regions of the reconstructed moiré superlattice.

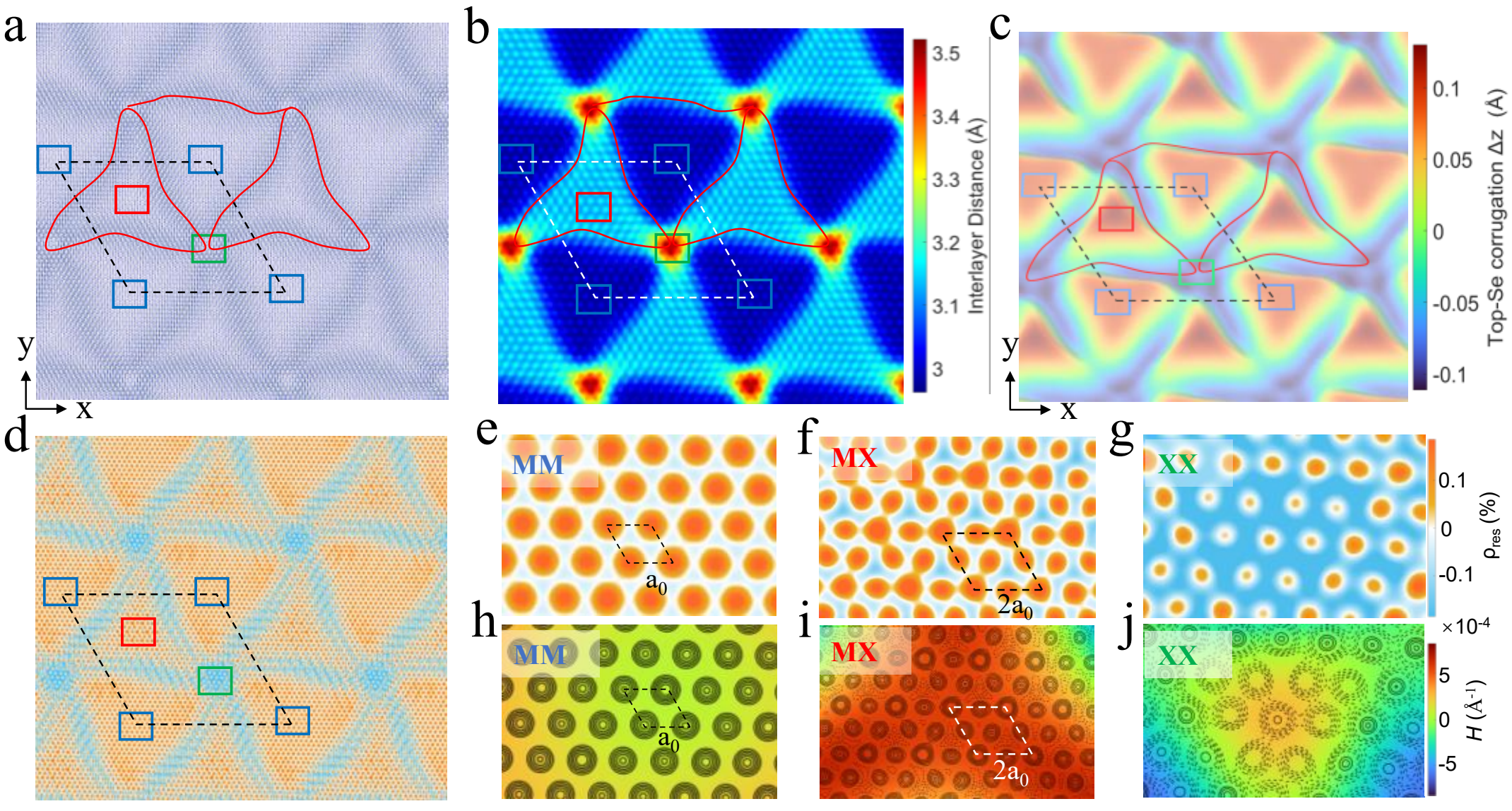

**Fig. 4 | Moiré-modulated lattice corrugation and curvature-coupled Ti atomic density in a 1.57° twisted $TiSe_2$/$NbSe_2$ heterostructure. a.** Top view of the relaxed $TiSe_2$/$NbSe_2$ moiré superlattice. **b.** Spatial map of the interlayer Se–Se distance. **c.** Relative height corrugation of the topmost Se layer. The dashed rhombus marks the moiré unit cell, and the solid red lines outline the characteristic vortex-like moiré pattern. The blue, red and green boxes indicate representative MM, MX and XX stacking regions, respectively. **d.** Smoothed map of Ti atomic-density residuals. **e–g.** Enlarged views of the Ti density modulation in the MM, MX, and XX stacking regions indicated in panel d. **h–j.** Curvature-resolved maps corresponding to the same stacking regions. The colored background represents the signed mean curvature ($-H$) of the smoothed Se surface, while the overlaid contour lines depict the residual Ti atomic density, with solid (dashed) black lines indicating positive (negative) values. The dashed diamond lines indicate the periodicity of the local configurations.

In two-dimensional van der Waals heterostructures, interlayer lattice mismatch and moiré/twist interactions relax the constraints of a perfectly rigid lattice, thereby providing a topological platform for engineering correlated electronic states. At present, the physical origin of the intrinsic CDW in monolayer 1T- $TiSe_2$ is widely considered to arise from the cooperative interplay between a gap-driven excitonic-insulator transition[34] and momentum-dependent electron–phonon coupling, such as Jahn–Teller-like distortion[35]. However, both mechanisms are generally understood within a framework that assumes a well-defined global electronic structure and long-range lattice translational symmetry. In our synthesized V-$TiSe_2$/2H-$NbSe_2$ heterostructure, the highly non-uniform interfacial stress imposed by the underlying 2H-$NbSe_2$ induces pronounced morphological distortion during structural relaxation governed by van der Waals interactions. This distortion not only produces a vortex-like non-Euclidean geometry extending over several nanometres, but also introduces local strain gradients on the microscopic scale[36].

According to continuum elasticity theory and two-dimensional Dirac physics, curved morphology and strain gradients in monolayer materials can effectively generate strong synthetic gauge fields and the associated pseudomagnetic fields[23, 24]. When combined with the strong interfacial electron-doping effect established above by STS and ARPES, this curvature-induced gauge field may drive a dramatic local reorganization of the electronic phase space in momentum space. Owing to the spatial redistribution of electron density and the localization of electronic wavefunctions by the pseudomagnetic field, the long-range phase coherence of the CDW—originally capable of extending across the entire micrometre-scale sample—is completely destroyed. This naturally explains why the nonlinear superstructure cannot form globally, but instead is excited only within specific morphological regions. More specifically, we attribute this localized and multi-oriented CDW state to a mechanism of strain-compensated local instability. As discussed above, extremely high electron doping would normally suppress the CDW phase of $TiSe_2$ altogether, as indeed observed in the relatively flat $1 \times 1$ regions elsewhere[37]. However, at the highly curved domain boundaries within the vortex structure, the extreme local tensile or compressive strain effectively modifies the local lattice constants and orbital overlap integrals, and thereby, within a confined region, partially compensates for the band mismatch caused by excessive electron doping[38]. As a result, the thermodynamic instability necessary for CDW formation can be locally restored only at these curved topological domain walls and in the neighbouring non-Euclidean regions, where the CDW nucleates and manifests as a geometrically constrained nonlinear state. Furthermore, because the domain walls themselves act as low-dimensional topological channels, the emergent CDW must extend one-dimensionally along the curved domain boundaries and become phase-pinned there[6]. This ultimately gives rise to multi-oriented superstructure features in the Fourier transform that are decoupled from the high-symmetry directions of the underlying lattice. The CDW observed on the V- $TiSe_2$ surface therefore does not originate from simple lattice defects or a uniform phase transition, but rather from an intense real-space competition and compromise among stress-induced topological geometric curvature, polarization-driven charge transfer, and many-body interactions. This mechanism offers a fundamentally new perspective for understanding the evolution of electronic states in correlated quantum materials on curved manifolds.

In summary, by constructing a monolayer 1T- $TiSe_2$/2H-$NbSe_2$ van der Waals heterostructure, we experimentally observe and systematically establish, for the first time, a new form of localized charge-density wave (CDW) existing in non-Euclidean space. Combining high-resolution real-space atomic imaging (STM/STS) with momentum-space electronic-structure characterization (ARPES), we uncover the complex microscopic competition among geometric curvature, strong interfacial charge transfer in this system. Crucially, our theoretical modeling reveals that the interplay between interfacial twist and lattice mismatch spontaneously generates a corrugated superlattice, where stacking-dependent interlayer coupling and local curvature synergistically dictate the spatial survival and geometric distortion of the CDW order. Our results show that non-Euclidean geometric distortion at the heterointerface not only induces strong pseudomagnetic fields and breaks the long-range translational symmetry of conventional CDWs in flat space, but also, through a unique strain-compensation effect, enables the spatial localization and multi-oriented phase pinning of correlated electronic order at topological domain boundaries. This finding fundamentally extends the conventional understanding that correlated electronic states in two-dimensional condensed-matter systems exist only in flat Euclidean space. More broadly, it establishes a profound coupling between geometric curvature and quantum many-body interactions, and provides a framework for simulating and exploring macroscopic quantum phenomena in non-Euclidean geometries.

**Acknowledgements**

We acknowledge financial support from the National Key R&D Program of China (Grants No. 2024YFA1207800), the National Natural Science Foundation of China (Grants Nos. U23A6015), CAS Project for Young Scientists in Basic Research (YSBR-003), the Fundamental Research Funds for the Central Universities. Calculations were performed at the Physics Lab of High-Performance Computing of Renmin University of China, Shanghai Supercomputer Center and Beijing Supercomputing Center.

**Competing interests**

The authors declare no competing interests.

**Note added:**

"During the preparation of this manuscript, we became aware of a recent study by Chunlei Gao *et al.* (*Nature Communications*) which investigates monolayer $TiSe_2$ on a bulk $NbSe_2$ substrate, unlike our monolayer $TiSe_2$/monolayer $NbSe_2$ heterostructure on BLG/SiC."

## Methods

### Experimental methods

Monolayer 1T-$TiSe_2$/2H-$NbSe_2$ heterostructure was successfully synthesized on the BLG/SiC substrate using a molecular beam epitaxy (MBE) system under an ultrahigh vacuum with a base pressure of less than $4\times10^{-10}$ mbar. Initially, the BLG on SiC(0001) was prepared by flashing to 1300°C a few times within an ultrahigh vacuum environment. Subsequently, monolayer 2H-$NbSe_2$ was synthesized directly onto the prepared BLG/SiC(0001) surface through the simultaneous evaporation of high-purity niobium (99.99%) and selenium (99.999%). During the deposition of niobium and selenium, the substrate temperature was maintained at approximately 400 °C, achieving a deposition rate of about 0.8 monolayers (ML) per half hour. Verification of surface integrity by STM (Fig.S1a), monolayer 2H-$NbSe_2$ was then utilized as the foundation for the vertical epitaxial growth of a monolayer 1T-$TiSe_2$. For the final phase, the monolayer 1T-$TiSe_2$ was fabricated by the co-evaporation of titanium (99.99%) and selenium (99.999%) onto the monolayer 2H-$NbSe_2$ surface, which was sustained at around 400°C. This growth process proceeded at a rate of approximately 1 ML per half hour. STM measurements were performed at constant-current mode in a base pressure environment of approximately $3 \times 10^{-10}$ mbar. ARPES measurements were taken at an even lower base pressure of approximately $6\times10^{-11}$ mbar, utilizing photon energies of 40.8 eV, with an energy resolution finer than 30 meV and an angular resolution of 0.3°. XPS measurements were performed with a hemispherical energy analyzer.

**Theoretical methods**

To relax the large-scale twisted $TiSe_2$/$NbSe_2$ moiré heterostructure, a pretrained Orb universal machine-learning interatomic potential was fine-tuned using system-specific density functional theory (DFT) reference data[1]. The DFT calculations were performed using the projector augmented wave method[2] and a plane-wave basis set, as implemented in the Vienna Ab-Initio Simulation Package (VASP)[3, 4]. The exchange-correlation potential was described by the r²SCAN meta-GGA functional[5], and long-range van der Waals interactions were included using Grimme's D3 dispersion correction with Becke–Johnson damping[6]. Kinetic energy cut-off of 700 eV and 400 eV were adopted for structural relaxations and ab initio molecular dynamics (AIMD) simulations, respectively. The Brillouin zone was sampled using a $3 \times 3 \times 1$ k-point mesh. The electronic convergence criterion was set to $1 \times 10^{-5}$ eV, and the force convergence threshold for the reference relaxations was $5 \times 10^{-3}$ eV/Å. The vacuum region of 15 Å was used to avoid the periodic interaction.

The reference dataset was constructed from representative $TiSe_2$/$NbSe_2$ configurations, including relaxed MM, MX and XX stacking structures calculated using $4 \times 4$ supercells, as well as AIMD trajectories sampled in the NVT ensemble at 300, 600 and 1000 K. In total, 4500 configurations were collected and labeled by DFT calculations. The DFT-labeled dataset was split into training, validation, and test subsets with a ratio of 9:0.5:0.5. The pretrained Orb model was fine-tuned for 100 epochs with a learning rate of $1 \times 10^{-7}$ and a batch size of 8. The model checkpoint with the lowest validation force error was used for subsequent structural relaxations. Model performance was evaluated on the held-out test set by comparing the predicted energies and forces with the corresponding DFT reference values (Fig. S2).

The fine-tuned Orb potential was then employed to relax the 1.57° twisted $TiSe_2/NbSe_2$ moiré superlattice. During the relaxation, the in-plane lattice vectors were fixed and all atomic coordinates were optimized until the maximum residual force was below $1\times10^{-2}$ eV/Å.

The in-plane Ti atomic positions in the $TiSe_2$ layer were converted into a continuous smeared atomic density, given by

$$\rho_{\mathrm{Ti}}(\mathbf{r}) = \sum_{i=1}^{N_{\mathrm{Ti}}} \frac{1}{2\pi\sigma^2} \mathrm{e}^{-|\mathbf{r}-\mathbf{R}_i|^2/(2\sigma^2)}$$

where $\mathbf{R}_i$ is the in-plane position of the $i$-th Ti atom and $N_{Ti}$ is the number of Ti atoms in the $TiSe_2$ layer. The Gaussian width was set to $\sigma = 0.55a_0$, where $a_0 = 3.545$ Å is the in-plane Ti–Ti lattice constant of monolayer $TiSe_2$. To remove the slowly varying moiré-scale density background, a long-wavelength density $\rho_{\mathrm{moiré}}(\mathbf{r})$ was obtained by Gaussian smoothing $\rho_{\mathrm{Ti}}(\mathbf{r})$ with a width of $1.5a_0$. The Ti atomic-density residual was defined as

$$\rho_{\mathrm{res}}(\mathbf{r}) = \frac{\rho_{\mathrm{Ti}}(\mathbf{r})-\rho_{\mathrm{moiré}}(\mathbf{r})}{\rho_{\mathrm{moiré}}(\mathbf{r})}\times 100$$

This normalization suppresses the moiré-scale stacking contrast and highlights the short-period Ti-density modulation associated with the CDW distortion.

The surface corrugation was extracted from the topmost Se layer of the relaxed heterostructure. The discrete Se atomic heights were converted into a continuous height field $h\ (x, y)$ on a periodic two-dimensional grid by Gaussian-weighted averaging,

$$h(\mathbf{r}) = \frac{\sum_j z_j e^{-|\mathbf{r}-\mathbf{R}_\mathrm{i}|/(2\sigma_h^2)}}{\sum_j e^{-|\mathbf{r}-\mathbf{R}_\mathrm{i}|/(2\sigma_h^2)}}$$

where $\mathbf{R}_j$ and $z_j$ are the in-plane position and height of the j-th topmost Se atom, respectively. The smoothing width was chosen as $\sigma_h = 1.2a_0$. The mean height was subtracted from $h(\mathbf{r})$ before the curvature analysis.

The local mean curvature of the height-field surface $z = h\ (x, y)$ was calculated as

$$H = \frac{(1+h_y^2)h_{xx} - 2h_x h_y h_{xy} + (1+h_x^2)h_{yy}}{2(1+h_x^2+h_y^2)^{3/2}}$$

where $h_x$, $h_y$, $h_{xx}$, $h_{yy}$, and $h_{xy}$ are the first and second spatial derivatives of the smoothed height field. In the curvature-resolved maps, the color background represents the signed mean curvature −H of the smoothed top-Se surface.

**Calculation of Curvature**

For Fig. 2d, we performed surface fitting and curvature calculations for the two regions. The blue contour corresponds to the region where the charge density wave (CDW) emerges. Its raw height profile, $z_{\mathrm{raw}}(x)$, contains local microscopic fluctuations superimposed on a continuous macroscopic envelope, $z_{\mathrm{env}}(x)$, which describes the elastic deformation of the 2D film. By performing a nonlinear Gaussian fit to extract the macroscopic envelope of the blue contour, we locate the deformation center at $x_c \approx 2.90$ nm. The peak height is approximately $H_{\mathrm{blue}} \approx 0.14$ nm with a full width at half maximum (FWHM) of $3.50$ nm. Based on the standard relationship $\mathrm{FWHM} = 2\sqrt{2\ln 2}\,\sigma$, the standard deviation is calculated to be $\sigma_{\mathrm{blue}} \approx 1.49$ nm. Consequently, the macroscopic surface envelope governing the non-Euclidean curvature in this region can be approximated as:

$$z_{\mathrm{blue_env}}(x) = 0.140\exp\left[-\frac{(x-2.90)^2}{4.44}\right]$$

where both $x$ and $z_{\mathrm{blue_env}}$ are in nanometers (nm).

Within the framework of curved-space continuum elasticity theory, synthetic gauge fields are highly sensitive to variations in the geometric curvature, $\kappa$. At the deformation center ($x = x_c$), the first derivative is zero, and the geometric curvature is determined by the absolute value of the second derivative:

$$\kappa_{\mathrm{apex}} = | z''_{\mathrm{env}}(x_c) | = \frac{H}{\sigma^2}$$

Substituting the fitted geometric parameters into this equation yields a macroscopic curvature at the apex of the blue region of:

$$\kappa_{\mathrm{blue}} \approx 0.0631\ \mathrm{nm}^{-1}$$

with a corresponding radius of curvature $R_{\mathrm{blue}} = 1/\kappa_{\mathrm{blue}} \approx 15.8$ nm.

Applying the exact same computational procedure to the gentler structural corrugations depicted by the yellow contour, we find a lower maximum envelope height of $H_{\mathrm{yellow}} \approx 0.115$ nm and a slightly broader FWHM of $3.65$ nm (corresponding to $\sigma_{\mathrm{yellow}} \approx 1.55$ nm). Applying the same curvature formula reveals a significantly smaller local curvature for the yellow region:

$$\kappa_{\text{yellow}} \approx 0.0479\ \text{nm}^{-1}$$

with a corresponding radius of curvature $R_{\text{yellow}} \approx 20.9$ nm.

A direct comparison of these geometric parameters ($\kappa_{\text{blue}} \approx 0.0631\ \text{nm}^{-1}$ versus $\kappa_{\text{yellow}} \approx 0.0479\ \text{nm}^{-1}$) clearly indicates that the blue region possesses a much larger geometric curvature. Such significant non-Euclidean geometric distortion in the α region (corresponding to the blue contour) induces stronger local strain gradients and larger pseudo-magnetic fields. These amplified geometric and topological effects overcome the elastic potential energy of the lattice, serving as the physical origin for the emergence and spatial localization of the novel CDW state; in contrast, this CDW state is absent in the gentler yellow region where the degree of bending is weaker.

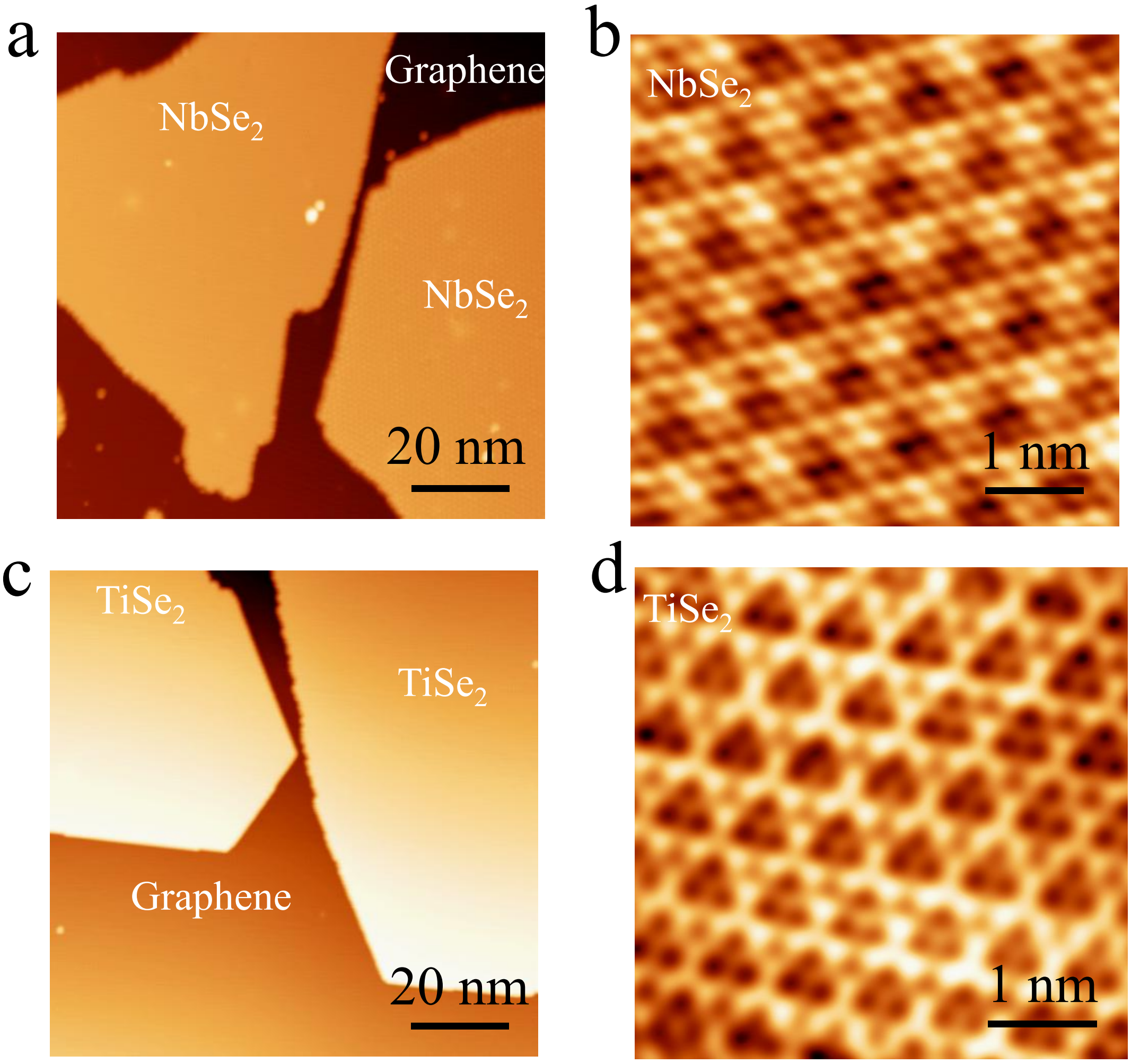

**Fig.S1 | STM images of monolayer $NbSe_2$ and $TiSe_2$ on a graphene surface. a, c** STM images of monolayer $NbSe_2$ (V=−2 V, I=200 pA) and $TiSe_2$ (V=2 V, I=100 pA) grown on graphene, respectively. **b, d** Atomic-resolution STM images of monolayer $NbSe_2$ (V=-0.5 V,

I=0.5 nA) and $TiSe_2$ (V=0.5 V, I=1.5 nA) in **a** and **c**, respectively.

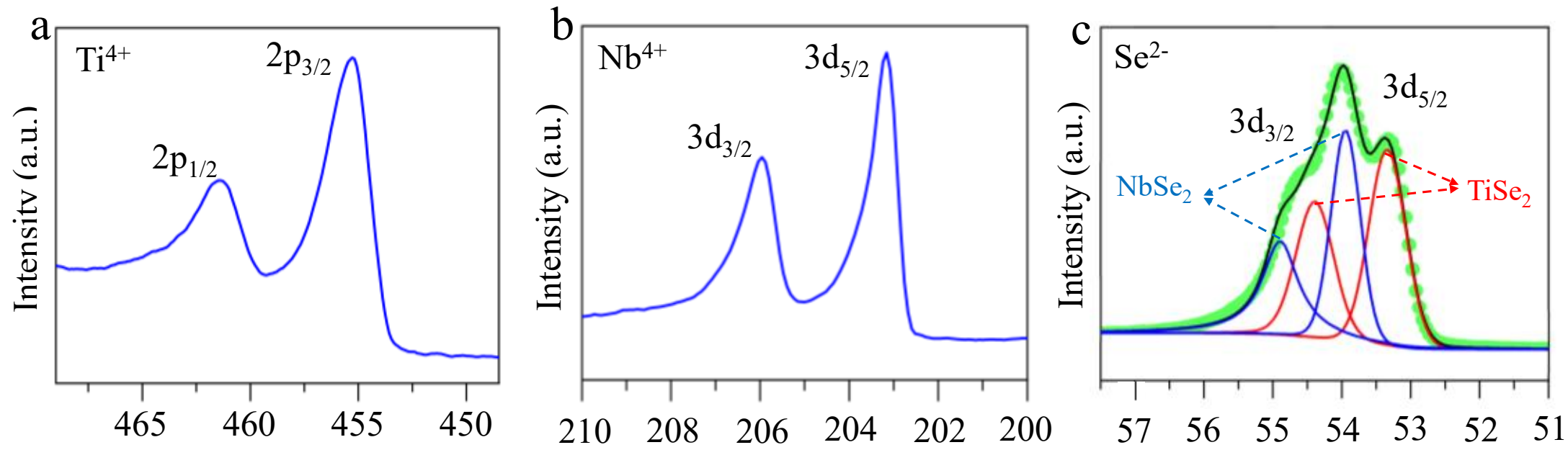

**Fig.S2 | XPS characterization of monolayer $TiSe_2$/$NbSe_2$ heterostructure. a, b** XPS measurements for the binding energies of Ti 2p and Nb 3d electrons, respectively. **c**. XPS spectra of the Se 3d core levels. The blue and red curves represent the fitting results corresponding to the Se 3d core levels of $NbSe_2$ and $TiSe_2$, respectively. The discrete points indicate the raw data, while the black line denotes the fitting curve.

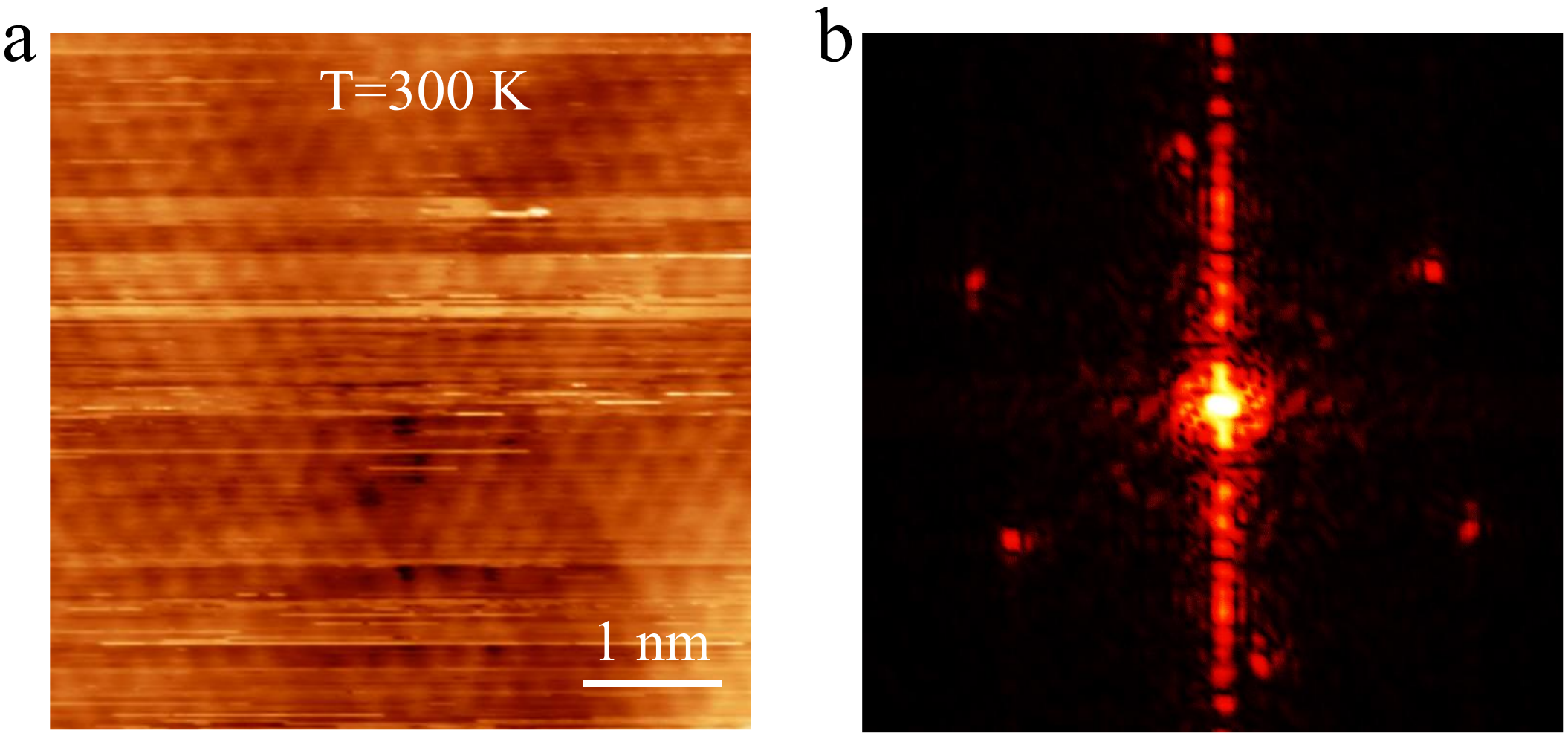

**Fig.S3 | STM characterization of monolayer V-$TiSe_2$ at 300 K. a.** STM image of monolayer V-$TiSe_2$ at 300 K (V=−100 mV, I=1 nA). **b** Fast Fourier transform (FFT) of STM image in **a**, showing the absence of superstructure.

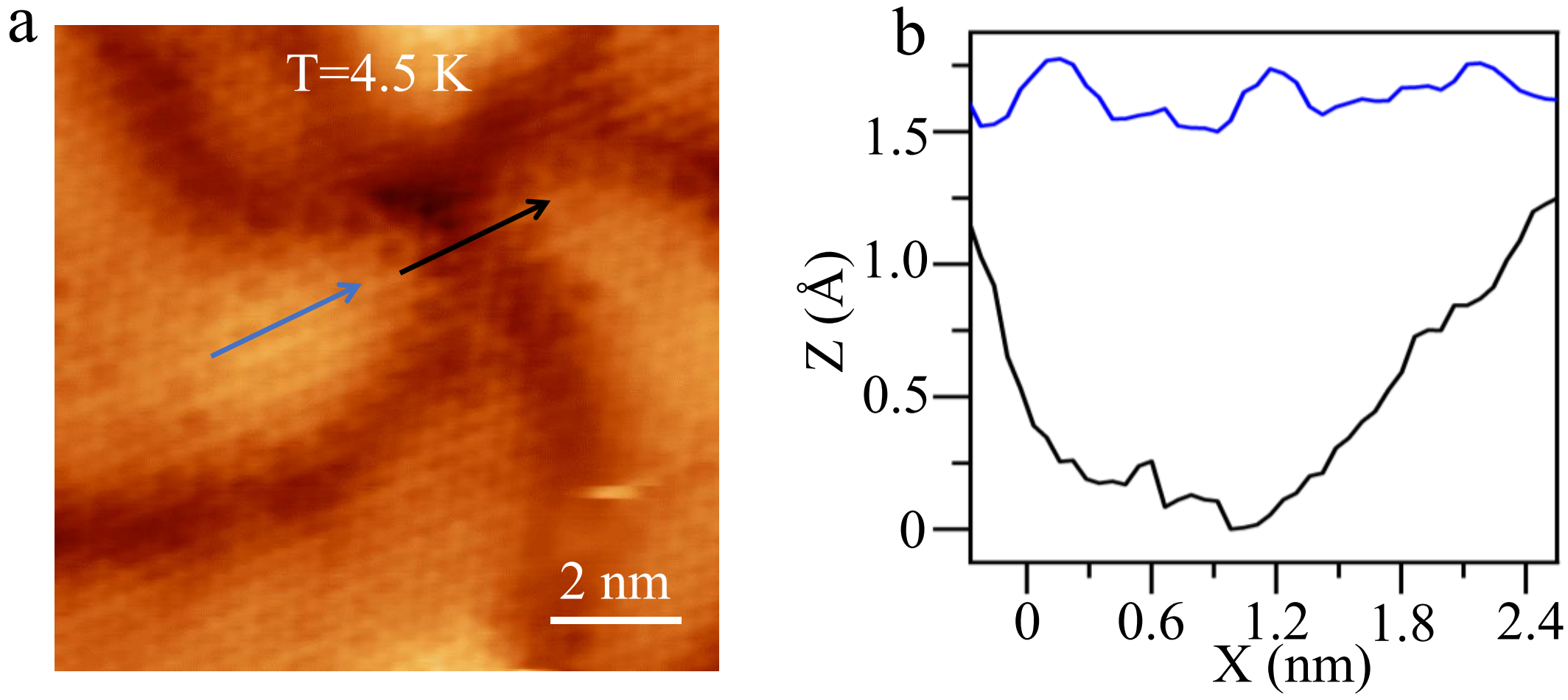

**Fig.S4 | Atomic corrugation of V-$TiSe_2$ and topographic variations at the domain boundary. a,** Monolayer V-$TiSe_2$ with vortex-like morphology at the heterostructure surface (V=−2 V, I=200 pA). **b,** Line profile taken along the blue and black arrows in a.

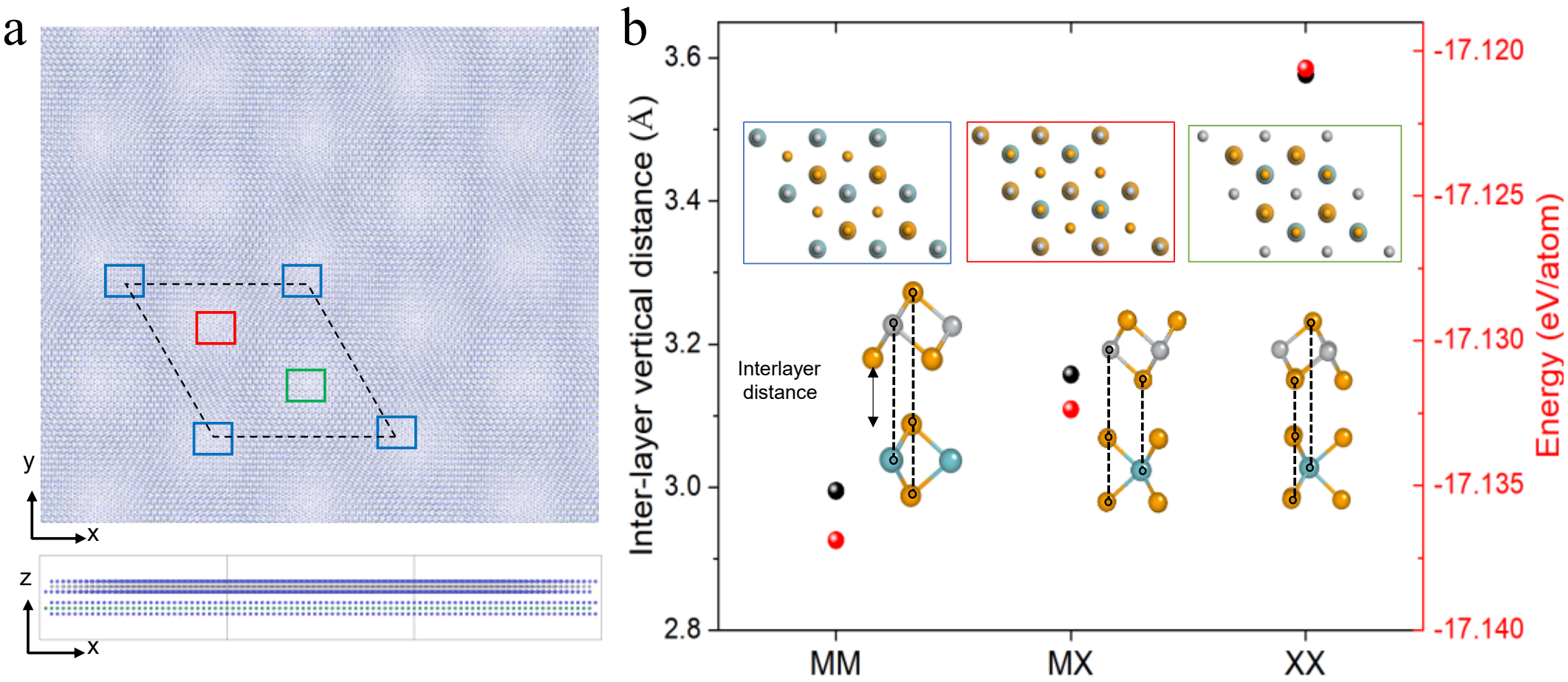

**Fig. S5 | Atomic model and stacking-dependent energetics of the twisted $TiSe_2$/$NbSe_2$ heterostructure. a**, structure of the unrelaxed $TiSe_2$/$NbSe_2$ heterostructure with a twist angle of 1.57°. **b.** Interlayer Se–Se distance and relative stacking energy for different local stacking configurations.

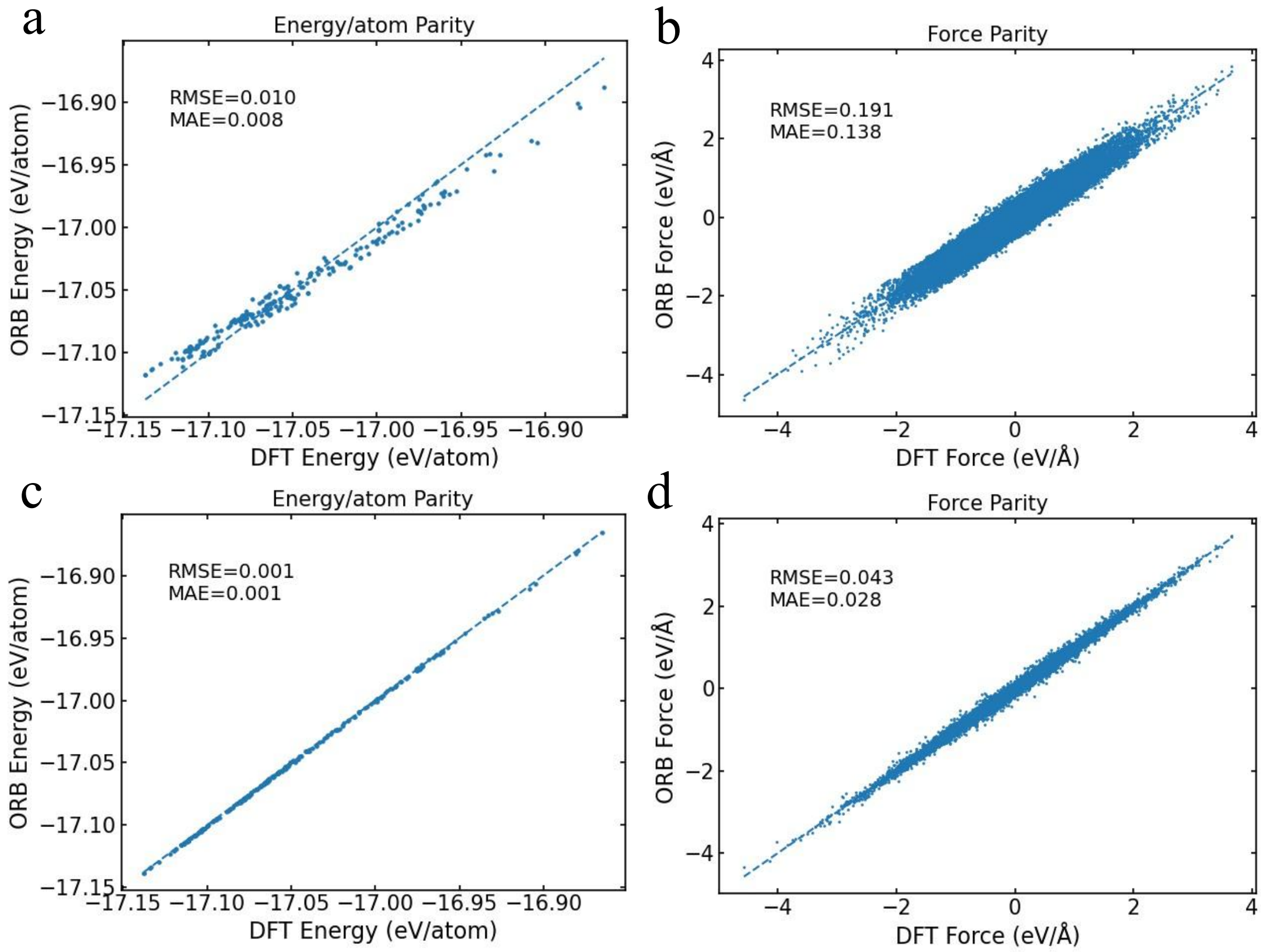

**Fig. S6. | Parity plots comparing Orb-model-predicted energies and forces with DFT reference values before and after fine-tuning. a, b.** Energy-per-atom and force parity plots obtained using the original Orb-model before fine-tuning. **c, d.** Corresponding energy-per-atom and force parity plots after Orb-model fine-tuning. The corresponding RMSE and MAE values are indicated in each panel. Before fine-tuning, the Orb model gave an energy RMSE/MAE of 0.010/0.008 eV/atom and a force RMSE/MAE of 0.191/0.138 eV/Å. After fine-tuning, these errors were reduced to 0.001/0.001 eV/atom and 0.043/0.028 eV/Å, respectively. The substantial reductions in RMSE and MAE demonstrate that fine-tuning significantly improves the accuracy of both energy and force predictions for the $TiSe_2$/$NbSe_2$ heterostructure system.